\documentclass[aps,prb,superscriptaddress,showpacs,twocolumn]{revtex4}
\usepackage{graphicx}
\begin{document}
\title{Flux Flow Instabilities in Microstructured Amorphous Nb$_{0.7}$Ge$_{0.3}$ Thin Films}
\author{D. Babi\'{c}}
\altaffiliation[Corresponding author] {} \email{dbabic@phy.hr}
\affiliation{Department of Physics, Faculty of Science, University of Zagreb,
Bijeni\v{c}ka 32, HR-10000 Zagreb, Croatia}
\author{J. Bentner}
\affiliation{Institut f\"{u}r experimentelle und angewandte Physik,
Universit\"{a}t  Regensburg, D-93025 Regensburg, Germany}
\author{C. S\"{u}rgers}
\affiliation{Physikalisches Institut, Universit\"{a}t Karlsruhe, D-76128 Karlsruhe, Germany}
\author{C. Strunk}
\affiliation{Institut f\"{u}r experimentelle und angewandte Physik,
Universit\"{a}t  Regensburg, D-93025 Regensburg, Germany}
%
%
%
%
\begin{abstract}
We report measurements of the electric field vs. current density [$E(J)$]
characteristics in the mixed state of amorphous Nb$_{0.7}$Ge$_{0.3}$ microbridges.
Close to the transition temperature $T_c$ the Larkin-Ovchinnikov theory of nonlinear
flux flow and the related instability describes the data quantitatively up to $\sim
70$ \% of the upper critical magnetic field $B_{c2}$ and over a wide electric field
range. At lower temperatures the nonlinearities of $E(J)$ can be described by
electron heating which reduces $B_{c2}$ and leads to a second type of flux flow
instability, as shown by a scaling analysis of the high-dissipation data.
\end{abstract}
%
%
\maketitle

It was predicted by Larkin and Ovchinnikov (LO) that the $E(J)$ curves of a "dirty"
superconductor in the mixed state may exhibit a steep increase long before the
depairing current density is reached.\cite{lo} This behaviour, called LO flux flow
instability (FFI),  appears when the energy of quasiparticles in driven vortex cores
becomes large enough for their escape into the surrounding superfluid. As a
consequence, the vortices shrink and the vortex motion viscosity is reduced,
resulting in an increase of the dissipation at the fixed $J$. According to LO, the
above mechanism is strong at temperatures $T$ close to $T_c$ and the corresponding
FFI occurs when $E$ reaches a critical value $E_i$ proportional to magnetic field
$B$, which implies a $B$-independent critical vortex velocity $u_i = E_i/B$. An
extension of the LO theory by Bezuglyj and Shklovskij (BS),\cite{bs} who took into
account heating effects due to a finite rate of heat removal to the bath, limited
the pure, nonthermal LO FFI to a $B$ considerably smaller than $B_{c2}$. In order to
explain a $B$-dependent $u_i$, other modifications of the LO theory explored
nonlocality of the spatial distribution of the excitations, caused by the
quasiparticle energy relaxation length being smaller than the intervortex
spacing.\cite{doet} Explanations of the FFI beyond the original or modified LO
picture were sought in dynamic vortex lattice crystallisation,\cite{kosh} depinning
phenomena,\cite{fend} appearance of hot spots,\cite{xiao} and recently in vortex
core expansion due to electron heating at low temperatures.\cite{kunchur}
Irrespective of the microscopic origin of the FFI, its distinct feature is an $E(J)$
region just above $E_i$ where theory predicts $dE/dJ < 0$, implying not only an
abrupt  jump but also a hysteresis in $E(J)$, as shown experimentally in
Ref.~\onlinecite{samoilov}. As $B$ is increased the jump disappears and
$E(J)$ is turned to a smooth nonhysteretic curve.

Previous analyses of the mechanisms that cause the FFI relied mostly on
identification of the jump at $E_i (J_i)$ and discussion of the $(B, T)$ dependences
of $E_i$, $J_i$ and other related parameters ($u_i$, power density $J_i E_i$, etc.)
The quantitative description of $E(J)$ over a wide $E$-range, i.e., extending both
below and above $E_i$, has remained an open question. Furthermore, $E_i$ is a
relevant parameter even if there is no jump but it has not been extracted from the
nonhysteretic $E(J)$.
 The lack of such an investigation in conventional
superconductors could possibly be explained by the usually strong pinning, which
complicates treatments of pure flux flow effects even in simple vortex systems. In
high-$T_c$ superconductors the pinning is weak, but the exact form of vortices is in
this case less well known because of their complex anisotropic character and
peculiar fluctuation phenomena in the depinned state.\cite{jrc} To avoid the
mentioned obstacles as much as possible we have chosen a material already proven to
be appropriate for studying the fundamental mechanisms of vortex dynamics, namely
amorphous Nb$_{0.7}$Ge$_{0.3}$ thin film of thickness comparable to the coherence
length $\xi$.\cite{basel} These samples have very weak or negligible pinning over a
considerable part of the $(B,T)$ plane and represent a simple classical "dirty"
superconductor with a well defined vortex structure. In addition, we have
reduced the measurement current and thus the power dissipation
in the sample by patterning strips of a few micrometre width.

Close to $T_c$ we have found a convincing quantitative agreement with the LO theory
up to an unexpectedly high $b = B/B_{c2} \sim 0.7$, in both the close-to-equilibrium
flux flow resistivity $\rho_f$ and the $E(J)$ extended over a wide range of
$J$. In particular, we show that $E_i$ can be determined unambiguously even if there
is no clearly defined jump. At low temperatures the LO description breaks down,
which suggests a different physics of the FFI. These data can be explained
consistently by electron heating to a temperature $T^*$ above the bath temperature
$T_0$. The electron heating causes a decrease of $B_{c2} (T^*)$ and eventually a
transition to the normal state at a certain electric field $E_c$.

The methods of sample fabrication and determination of superconducting parameters are described in
Ref.~\onlinecite{basel}. The measured microbridge,\cite{sample2} deposited onto an oxidised Si substrate,
is 210 $\mu$m long, 5 $\mu$m wide, 20 nm thick,
and has the following parameters of interest:
$T_c =$ 2.75 K (the transition width of 0.05 K),
the estimated $T = 0$ normal state resistivity $\rho_n (0) = 3.3 \pm 0.2$ $\mu \Omega$m,
$ - (dB_{c2} / dT)_{T=T_c} \approx 2.6$ TK$^{-1}$,
$\xi(0) = 6.8$ nm, and the other Ginzburg-Landau parameters are
 $\kappa$ = 103 and
$\lambda(0) = 1.15$ $\mu$m. All the sample parameters are within the range of
expected values for amorphous Nb$_{0.7}$Ge$_{0.3}$ thin films. The measurements were
performed in a $^3$He cryostat with rf filtered leads. The dc $E(J)$ was measured by
sweeping an applied current at a rate \mbox{10 nAs$^{-1}$} \mbox{(0.1
MAm$^{-2}$s$^{-1}$)}, whereas the magnetoresistivity [$\rho(B,T)$] measurements were
carried out using small currents \mbox{(1 MA/m$^2$)} at which the $E(J)$ is linear,
originating  from either thermally activated vortex hopping or free flux
flow.\cite{basel}

In Fig.~1 we show the $E(J)$ at $T_0 =2.5$ K ($t = T_0/T_c = 0.91$) for 0.1 T $\leq B \leq$
0.5 T ( $0.15 \leq b \leq  0.77$). All the curves were measured by increasing the
applied current. A change from an $E(J)$ with the FFI jump (low $B$) to a smooth
$E(J)$ (high $B$) is clearly visible, as well as a gradual approaching the normal
state electric field $E_n=  \rho_n J$  (dashed line)  at large $J$. We show below
that the LO FFI theory explains quantitatively all these curves. Close to $T_c$ the
LO expression for $J(E)$ is given by

\begin{equation}
\label{loej}
J = \sigma_n \left[ A + \frac{g(b)}{b(1-t)^{1/2}} Y(E) \right] E   \;  \;  \; ,
\end{equation}
where $\sigma_n = 1/ \rho_n$, $A$ is a constant of order unity, $Y(E)= 1/(1
+E^2/E_i^2)$ describes the vortex core shrinking,  and $g(b)$ is a function
approximated by the following interpolation formulae: $g_1(b) = 4.04 - b^{1/4} (3.96
+ 2.38b)$ for $b < 0.315$, and $g_2(b) = 0.43 (1 - b)^{3/2} + 0.69 (1 - b)^{5/2}$
for $b  >  0.315$. In the limit $E \ll E_i$, $Y (E) \approx 1$ and Eq.~\ref{loej}
gives the flux flow resistivity $\rho_f = E/J$. In the expression for $\rho_f$, $A=1$
follows from the condition $\rho_f (B_{c2}) = \rho_n$, whereas in nonequilibrium the
constant value of $A \approx 1$ reflects suppression of the superconducting order
parameter outside the cores by a strong electric field.\cite{samoilov}

%
\begin{figure}
\includegraphics[width=75mm]{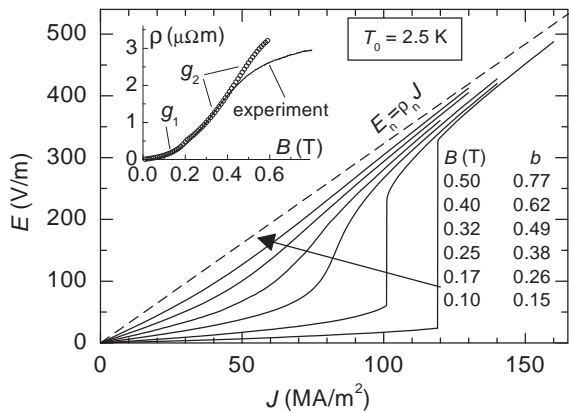}
\caption{$E(J)$ at $T_0=2.5$ K, for 0.1 T $\leq B \leq$ 0.5 T ($B_{c2} = 0.65 \pm
0.03$ T) increasing as indicated by the arrow, measured by sweeping the applied
current up. The dashed line represents $E_n=\rho_n J$. Inset: Measured
magnetoresistivity (solid line) and the LO $\rho_f$ (open symbols) plotted using
$g_1$ and $g_2$ as explained in the text.}
\end{figure}

A comparison of the measured data (full lines) and Eq.~\ref{loej} (dashed lines) is
shown in Fig.~2 for two characteristic shapes of  the $E(J)$ at $T_0 = 2.5$ K, i.e.,
for those with ($B = 0.1$ T) and without ($B = 0.4$ T) the jump. Eq.~\ref{loej}
agrees with the data excellently by taking $\sigma_n = 3.1 \times 10^5$ S/m from
$\rho (B \sim 2B_{c2})$, $A$ ranging from 0.92 to 0.97 with no systematic
$B$-dependence, and using $B_{c2} = 0.65 \pm 0.03 \rm{~T}$ to calculate $g(b)/b$ and
the corresponding error bars (important at low $b$ where $g(b)/b$ is a steep
function). Thus, the only free parameter left is $E_i$, shown in the inset to
Fig.~2(a)  and discussed later.

%
%
\begin{figure}
\includegraphics[width=75mm]{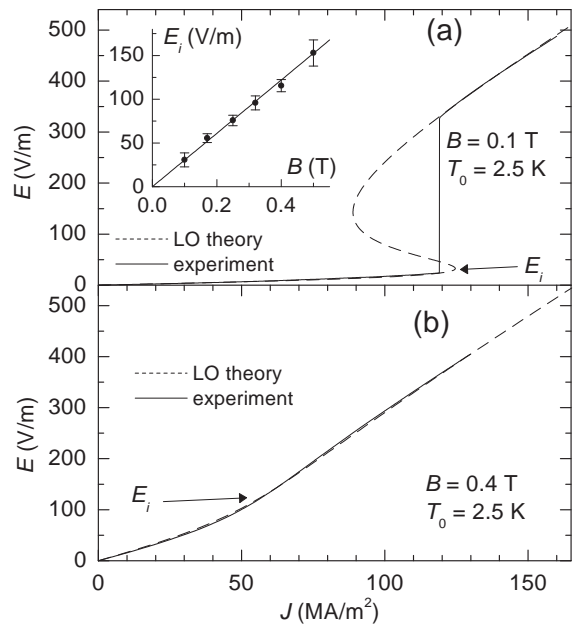}
\caption{$E(J)$ at $T_0 = 2.5$ K (full lines) measured at an increasing applied current,
for $B=0.1$ T (a) and $B=0.4$ T (b). The dashed lines are plots of Eq.~\ref{loej}
with the appropriate choices of the parameters, as discussed in the text. Inset to
(a):  Extracted $E_i(B)$ (circles), illustrating the validity of the LO theory with
a critical velocity $u_i=E_i/B$ independent of magnetic field.}
\end{figure}

In the inset to Fig.~1 we show $\rho (B)$ (solid line) at the same temperature,
compared with the theoretical $\rho_f$ (open symbols). With the same values of
parameters $B_{c2}$, $\sigma_n$ and $A$ as above, the agreement of the data and the
LO theory is satisfactory below $\sim 0.5$ T all the way down to  $B \rightarrow 0$.
This implies a negligible critical current density $J_c$ and a good description of
the close-to-equilibrium  transport properties in terms of the LO theory for all the
$E(J)$ shown in Fig.~1. The LO theory however fails to explain the data closer to
$B_{c2}$, in contrast to our previous finding\cite{basel} for another sample at $t =
0.82$.  To resolve this apparent inconsistency we measured $\rho(B)$ of the present
sample at $T_0 = 2$ K ($t = 0.7$) and found that the agreement of the data close to
$B_{c2}$ and the LO theory is restored at this temperature. The failure of the LO
theory to describe $\rho(B \rightarrow B_{c2})$ in the vicinity of $T_c$ may be
related to a widening of the {\it equilibrium} critical-fluctuation region at $B$
sufficiently close to $B_{c2}$.

From the slope of linear $E_i (B)$  we calculate the critical vortex velocity $u_i =
305$ m/s, the quasiparticle energy relaxation time\cite{lo} $\tau_\epsilon = D [14
\zeta(3) (1-t)]^{1/2}/ \pi u_i^2 = 0.18$ ns near $T_c$, where $\zeta$ is the Riemann
zeta function, $D = 8 k_B T_c \xi^2(0) / \pi \hbar = 4.3 \times 10^{-5}$ m$^2$/s the
diffusion constant, and the corresponding inelastic relaxation length $l_\epsilon =
\sqrt{D \tau_\epsilon} = 87$ nm. The linearity of $E_i(B)$ provides strong evidence
for the FFI being caused by the LO mechanism of vortex core shrinking. Note that the
LO model holds up to an unexpectedly high $b$, which is twice larger than the upper
limit estimated by BS. Only for $B=0.5$ T the relatively large error bar of the
corresponding $E_i$ may imply that the BS heating is starting to take place, but
even there the agreement with Eq.~\ref{loej} is very good over the whole $E$-range.
Previously we have shown that the weak heating effects in this regime have
contributed mostly to the vortex motion noise.\cite{basel} In conclusion to this
part, our results for $T_0$ close to $T_c$ are over a large $B$-interval in
remarkable quantitative agreement with the LO theory.

We now turn to the low temperature regime. Recently Kunchur analysed the FFI in
YBa$_2$Cu$_3$O$_{7-\delta}$ at low temperatures and small-to-moderate $b$ in terms
of electron heating to a temperature $T^*(E) > T_0$. The heating occurs due to
insufficient heat transfer to phonons at low temperatures, which reduces the
superconducting order parameter and results in a decrease of $B_{c2}
(T^*)$.\cite{kunchur} Having found the above excellent agreement with the LO picture
close to $T_c$, this urged us to carry out low-$T_0$ measurements over a similar
range of $b$ and thus investigate the differences and/or similarities between the
FFI and overall nonlinearities of the $E(J)$ at low an high temperatures. The
resulting set of $E(J)$ measurements at $T_0=1.1$ K ($t=0.4$) is shown in Fig.~3. To
check whether the LO approach could be valid thus below $T_c$ we tried to modify
Eq.~1 by using the given form of $Y(E)$ and replacing the $b$-dependent part with
the ones appropriate at low temperatures (see Eqs.~\ref{flow},\ref{fhigh} below).
Despite the apparent similarity of the curves when compared to those of Fig.~2, we
did not obtain any satisfactory agreement even if all the numerical parameters were
left floating. This motivated us to ana\-lyse these results in terms of electron
heating as the cause of a second type of the FFI.
%
%
\begin{figure}
\includegraphics[width=75mm]{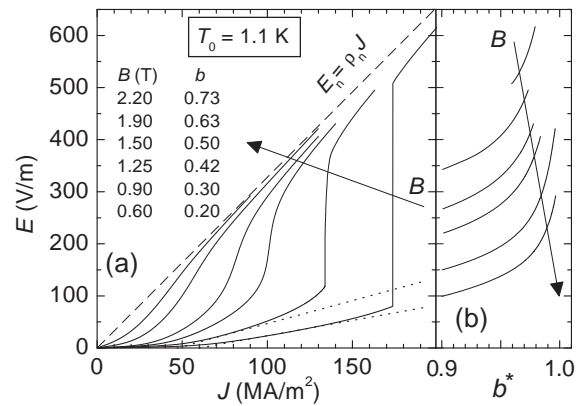}
\caption{(a) $E(J)$ at $T_0=1.1$ K (solid lines), for 0.6 T $\leq B \leq$ 2.2 T
($B_{c2} = 3.0 \pm 0.1$ T) increasing as indicated by the arrow. The measurements
correspond to an ascending applied current. The dashed line shows $E_n = \rho_n J$.
The dotted lines are plots of Eq.~\ref{flow} for 0.6 T and 0.9 T. (b) $E$ vs. $b^*$
calculated from the measured $E(J)$ and Eq.~\ref{fhigh} using $\alpha=3$. 
The vertical scale is the same as in (a) and the arrow 
points again in the direction of increasing $B$.}
\end{figure}

At low $t$ and $b$ the expression for $J(E)$ is given by\cite{lo}

\begin{equation}
\label{flow} J = J_c + \frac{\sigma_n}{0.9\;b} E \; \; \; .
\end{equation}
Eq.~\ref{flow} predicts that if the heating is weak  the slope of $E(J)$
sufficiently above $J_c$ is $0.9 \rho_n b$. The dotted lines in Fig.~3 show  plots
of Eq.~\ref{flow} for 0.6 T and 0.9 T, where the linear region can be found with
reasonable reliability. We used $B_{c2} = 3.0 \pm 0.1$ T as determined from $\rho_f
(B)$ and $J_c$ being the only free parameter. A replacement  $T_0 \rightarrow
T^*(E)$ in Eq.~\ref{flow} can explain the rise of $E(J)$ above the dotted lines
by a progressive decrease of $J_c$ and increase of $b$. Moreover, the ratio $E/b(E)$ may
depend nonmonotonically on $E$. This results in a negative slope of $J(E)$ and thus
causes another type of flux flow instability. However, Eq.~\ref{flow} does not
describe the $E(J)$ at large $E$, since the upper limit of the dissipation (given by
$b =1$) cannot explain the data close to $E_n$. This topic was not addressed by
Kunchur, who concentrated only on the low-$b$ and low-$E$ regime.

In order to analyse the $E(J) \rightarrow E_n$ data
we recall another LO result, namely that as long as the electron mean free path
is much smaller than $\xi$, close to $B_{c2}$ the $J(E)$ is at an
arbitrary temperature determined by\cite{lo}

\begin{equation}
\label{fhigh}
J = \sigma_n \left[ 1 + \alpha(T)\;(1 - b)\right] E  \; \; \; ,
\end{equation}
where $\alpha$ is a temperature-dependent constant varying between 2 and 4, and
$J_c$ at such high dissipation can be disregarded. The above expression was used
successfully in a detailed analysis of the high-$b$ magnetoresistance of amorphous
Nb$_{0.7}$Ge$_{0,3}$ films.\cite{geers} If the assumption of
electron-heating-induced nonlinearities is correct, Eq.~\ref{fhigh} should describe
the upper part of $E(J)$ through $E$-dependence of $b$ and $\alpha$ up to the
transition to the normal state at a critical electric field $E_c (B)$ corresponding
to $T^* = T_c (B)$ (equivalently, to $B = B_{c2} (T^*)$). In Eq.~\ref{fhigh} the
heating affects $\alpha$ and increases $b$ to a {\it nonequilibrium} $b^* [ T^*
(E)]$. The temperature dependence (and hence the $E$-dependence) of $\alpha$ is
weak,\cite{lo} and if we approximate $\alpha$ by a constant we can invert
Eq.~\ref{fhigh} to calculate $b^*(E) = B/B_{c2}(T^*)$ from our $E(J)$ data. In
Fig.~3(b) we show a plot of $E$ vs. $b^*$ (calculated using $\alpha = 3$) for $b^* >
0.9$, where we expect the validity of Eq.~\ref{fhigh} and the approximation of a
constant $\alpha$.

Since the heating characteristics $T^*(E)$ is mainly governed by electron-phonon
scattering it should be independent of $B$, the effect of which is expressed through
the $E_c (B)$ dependence. Thus, we expect a {\it scaling behaviour} of $J(E)$ for
different magnetic fields provided $E_c (B)$ is taken into account. This is
demonstrated in Fig.~4, where $E_c(B)$ is chosen such that $b^*$ (calculated with
$\alpha=3$) scales with $1-E/E_c(B)$, i.e., the data shown in Fig.~3(b) can be
collapsed onto the same curve. The corresponding electron temperature is 1.7 K $<
T^* <$ 2.5 K, as estimated from the equilibrium $B_{c2} (T)$ characteristics. In the
inset to Fig.~4 we plot $E_c$ against equilibrium $1 - b$. The error bars indicate
the range of variations of $E_c$ when $\alpha$ changes between 2 to 4 (and results
in a satisfactory scaling).

%
\begin{figure}
\includegraphics[width=70mm]{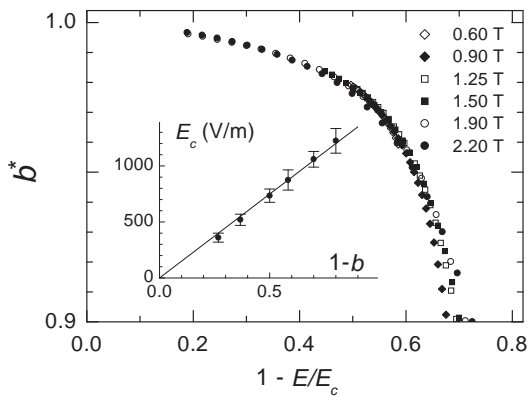}
\caption{Scaling plot of the  nonequilibrium reduced magnetic field $b^*$ vs. $1 -
E/E_c$ calculated from Eq.~\ref{fhigh} using $\alpha = 3$, where $E_c(B)$ is plotted
against equilibrium $(1 - b)$ in the inset. The error bars of $E_c$ indicate how
much $E_c$ varies if the scaling is performed using $\alpha$ between 2 and 4.}
\end{figure}

The above procedure corresponds to a determination of $E_c(B)$. The solid line in
the inset to Fig.~4 represents $E_c = E_{c0} (1-b)$, where $E_{c0} = 1500$ V/m, and
describes the inferred values of $E_c$ fairly well. This $E_c (b)$ dependence can be
related to the thermodynamic properties of the mixed state. For a "dirty"
high-$\kappa$ superconductor the Gibbs free energy density stored in the superfluid
at sufficiently large $b$ is given by $G_s = U_c (1-b)^2$, where $U_c = B_{c2}^2 / 4
\mu_0 \kappa^2$ is the zero-$B$ superconducting condensation energy.\cite{fetter} In
the present case, $U_c (1.1 \; {\rm K}) \approx 170$ J/m$^3$. Assuming that electron
heating weakens the superfluid we can equate $G_s = \sigma_n E_c^2 \tau_\epsilon$,
which explains $E_c \propto 1 - b$ provided $\tau_\epsilon$ does not depend on $B$
(expected for electron-phonon scattering). Furthermore, if $\tau_\epsilon$ is
considered as fairly a constant in the given $T^*$-range we can calculate
$\tau_\epsilon (T \rightarrow T_c) \approx U_c/\sigma_n E_{c0}^2 = 0.24$ ns, which
is in reasonable agreement with the independent estimate
$\tau_\epsilon=0.18\rm{~ns}$ from the $E(J)$ at $T_0 = 2.5$ K.

Our and Kunchur's complementary approaches agree in the final conclusion that
electron heating and the corresponding reduction of $B_{c2}$  is the main source of
nonlinearities of $E(J)$ in strongly nonequilibrium vortex transport at low
temperatures. Another implication of our analysis is that critical fluctuations
around $B_{c2}(T_0 \rightarrow T_c)$, which affect $\rho_f$ in complete thermal
equilibrium (see the inset to Fig.~1), seem to be of little relevance when $T_0 \ll
T_c$ even if $B \rightarrow B_{c2}$ by electron heating. This issue however requires
a more detailed investigation.

In conclusion, we have measured and analysed the $E(J)$ curves of amorphous
Nb$_{0.7}$Ge$_{0.3}$ microbridges over a wide range of magnetic field and in two
characteristic regimes, i.e., close to and well below $T_c$. In the former case we
have found an excellent agreement with the Larkin-Ovchinnikov theory of
nonlinear flux flow and the related
instability up to a surprisingly high value of $B/B_{c2} \sim 0.7$,
much larger than predicted theoretically by Bezuglyj and Shklovskiij.  At low
temperatures the nonlinearity of $E(J)$ and the
flux flow instability can be reasonably well described by electron
heating and the related decrease of $B_{c2}$. Our scaling analysis of the
$E(J)$ curves at high currents supports this conclusion quantitatively,
leading to an agreement with the thermodynamic properties of the
mixed state.

We thank F.~Rohlfing, W. Meindl, B.~Stojetz, A.~Bauer and M.~Furthmeier for
technical support. This work was partially funded by the Deutsche
Forschungsgemeinschaft within the Graduiertenkolleg 638. Additional support by the
Croatian Ministry of Science (project No. 119262) and the Bavarian Ministry for
Science, Research and Art is greatfully acknowledged.


\begin{thebibliography}{99}
\bibitem{sample2} All the $E(J)$ measurements presented were also carried out on another sample that contained a line
        of nanostructured holes, which gave a correction to the $E(J)$ of around 10 \% but did not alter the overall behaviour.
        These results will be published separately.
\end{thebibliography}
\end{document}